\documentclass[aps,prb,twocolumn,a4paper,floatfix,superscriptaddress,showpacs,showkeys]{revtex4-1}

\usepackage{dcolumn,amsmath,xspace}
\usepackage{graphicx}

\usepackage{epstopdf}

\newcommand{\degC}{\ensuremath{~^{\circ}\text{C }}}
\newcommand{\sixrt}{\ensuremath{(6\sqrt{3}\!\times\!6\sqrt{3})\text{R}30^\circ}~}

\begin{document}
\title{A wide band gap metal-semiconductor-metal nanostructure made entirely from graphene}

\author{J. Hicks}
\affiliation{The Georgia Institute of Technology, Atlanta, Georgia 30332-0430, USA}
\author{A. Tejeda}
\affiliation{Institut Jean Lamour, CNRS - Univ. de Nancy - UPV-Metz, 54506 Vandoeuvre les Nancy, France}
\affiliation{Synchrotron SOLEIL, L'Orme des Merisiers, Saint-Aubin, 91192 Gif sur Yvette, France}
\author{A. Taleb-Ibrahimi}
\affiliation{UR1 CNRS/Synchrotron SOLEIL, Saint-Aubin, 91192 Gif sur Yvette, France}
\author{M.S. Nevius}
\author{F. Wang}
\author{K. Shepperd}
\author{J. Palmer}
\affiliation{The Georgia Institute of Technology, Atlanta, Georgia 30332-0430, USA}
\author{F. Bertran}
\author{P. Le F\`{e}vre}
\affiliation{Synchrotron SOLEIL, L'Orme des Merisiers, Saint-Aubin, 91192 Gif sur Yvette, France}
\author{J. Kunc}
\author{W.A. de Heer}
\affiliation{The Georgia Institute of Technology, Atlanta, Georgia 30332-0430, USA}
\author{C. Berger}
\affiliation{The Georgia Institute of Technology, Atlanta, Georgia 30332-0430, USA}
\affiliation{CNRS/Institut N\'{e}el, BP166, 38042 Grenoble, France}
\author{E.H. Conrad}\email[email: ]{edward.conrad@physics.gatech.edu}
\affiliation{The Georgia Institute of Technology, Atlanta, Georgia 30332-0430, USA}

\begin{abstract}
\textbf{A blueprint for producing scalable digital graphene electronics has remained elusive. Current methods to produce semiconducting-metallic graphene networks all suffer from either stringent lithographic demands that prevent reproducibility, process-induced disorder in the graphene, or scalability issues. Using angle resolved photoemission, we have discovered a unique one-dimensional metallic-semiconducting-metallic junction made entirely from graphene, and produced without chemical functionalization or finite size patterning. 
The junction is produced by taking advantage of the inherent, atomically ordered, substrate-graphene interaction when it is grown on SiC, in this case when graphene is forced to grow over patterned SiC steps. This scalable bottom-up approach allows us to produce a semiconducting graphene strip whose width is precisely defined within a few graphene lattice constants, a level of precision entirely outside modern lithographic limits. The architecture demonstrated in this work is so robust that variations in the average electronic band structure of thousands of these patterned ribbons have little variation over length scales tens of microns long. The semiconducting graphene has a topologically defined few nanometer wide region with an energy gap greater than 0.5 eV in an otherwise continuous metallic graphene sheet. This work demonstrates how the graphene-substrate interaction can be used as a powerful tool to scalably modify graphene's electronic structure and opens a new direction in graphene electronics research.}
\end{abstract}
\vspace*{4ex}

\maketitle
\newpage

Patterning a flat graphene sheet to alter its electronic structure was envisaged to be the foundation of graphene electronics.\cite{Berger_JPC_04}  The early focus was to open a finite-size gap in lithographically patterned nanoribbons, a necessary step for digital electronics.\cite{Berger_JPC_04,Nakada_PRB_96,Wakabayashi_PRB_99,brey2006,Son_PRL_06} While early transport measurements supported this possibility,\cite{Melinda_PRL_07} it soon became apparent that these ``transport gaps'' originated from a series of mismatched-level quantum dots caused by the inability of current lithographically to produce sufficiently narrow, well ordered, and crystallography define graphene edges.\cite{Melinda_PRL_10,jiao2009,jiao2010,Oostinga_PRB_10}  A working solution to the graphene "gap problem" has yet to be formulated, let alone demonstrated. We show that in fact such a solution exists, not by patterning graphene, but instead by controlling the graphene-substrate geometry. 

We have been able to construct a unique, reproducible, and scalable semiconducting graphene ribbon with a gap larger than 0.5 eV. Using pre-patterned SiC trenches to force graphene to bend between a high symmetry (0001) face to a low symmetry $(11\bar{2}n)$ facet, we produce a narrow curved graphene bend with localized strain. This "topologically-defined" ribbon is a wide-gap graphene semiconductor strip a few lattice constants wide that extends hundreds of microns long.  The strip is connected seamlessly to metallic graphene sheets on both of its sides. One metallic sheet is n-doped and the other p-doped. From this simple morphology, we have not only produced a gap suitable for room temperature electronics, we have demonstrated how narrow Schottky barriers can be fabricated in a device-scalable architecture.  


\begin{figure*}[htb]
\includegraphics[angle=0,width=17.0cm,clip]{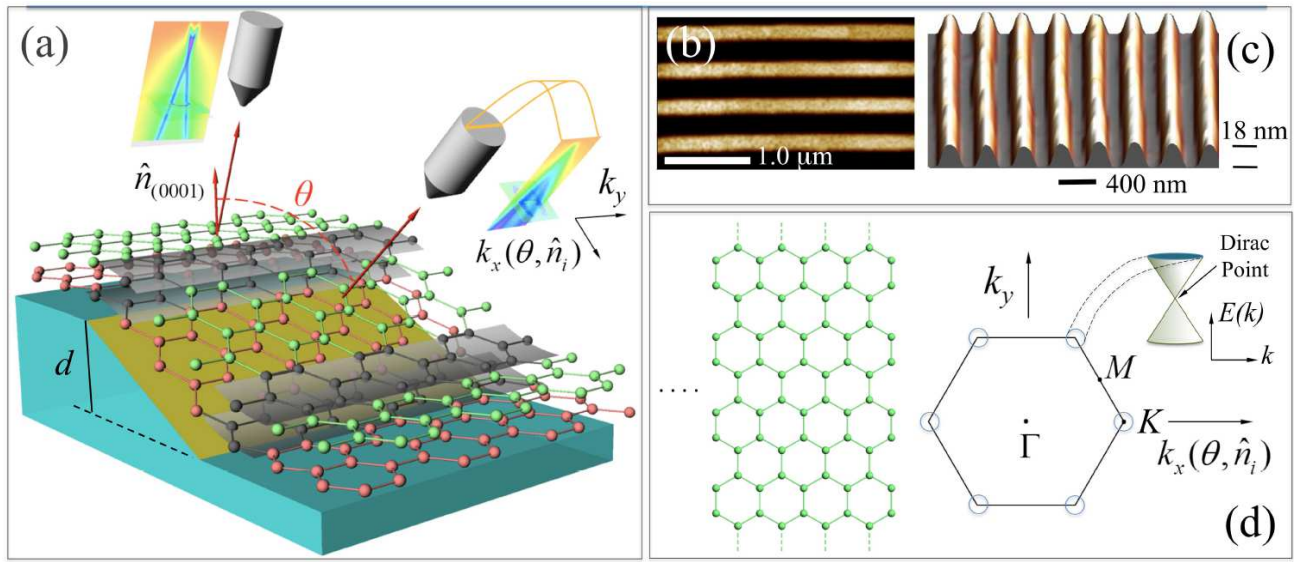}
\caption{(a) Schematic side view of the side wall ribbons containing two structural graphene sheets (the lower sheet in red is referred to as the ``buffer'' layer).  Grey regions are curved parts of the graphene that are semiconducting.  The schematic shows the detector orientation to reach both the $K$-point of the flat surface and the facet with normal $\hat{n}_i$. Sample $k_y$ and $k_x(\theta,\hat{n}_i)$ cuts through the $K$-points of the (0001) and facet surface show the Dirac cones and constant energy contours. (b) AFM top view of side wall grown ribbons showing the long range order. (c) Perspective AFM view of 18 nm deep graphitized trenches with a trench pitch of 400 nm. (d) The graphene BZ orientation relative to the armchair edge ribbons. }\label{Ribbons}
\end{figure*}

\begin{figure*}
\includegraphics[angle=0,width=15.0cm,clip]{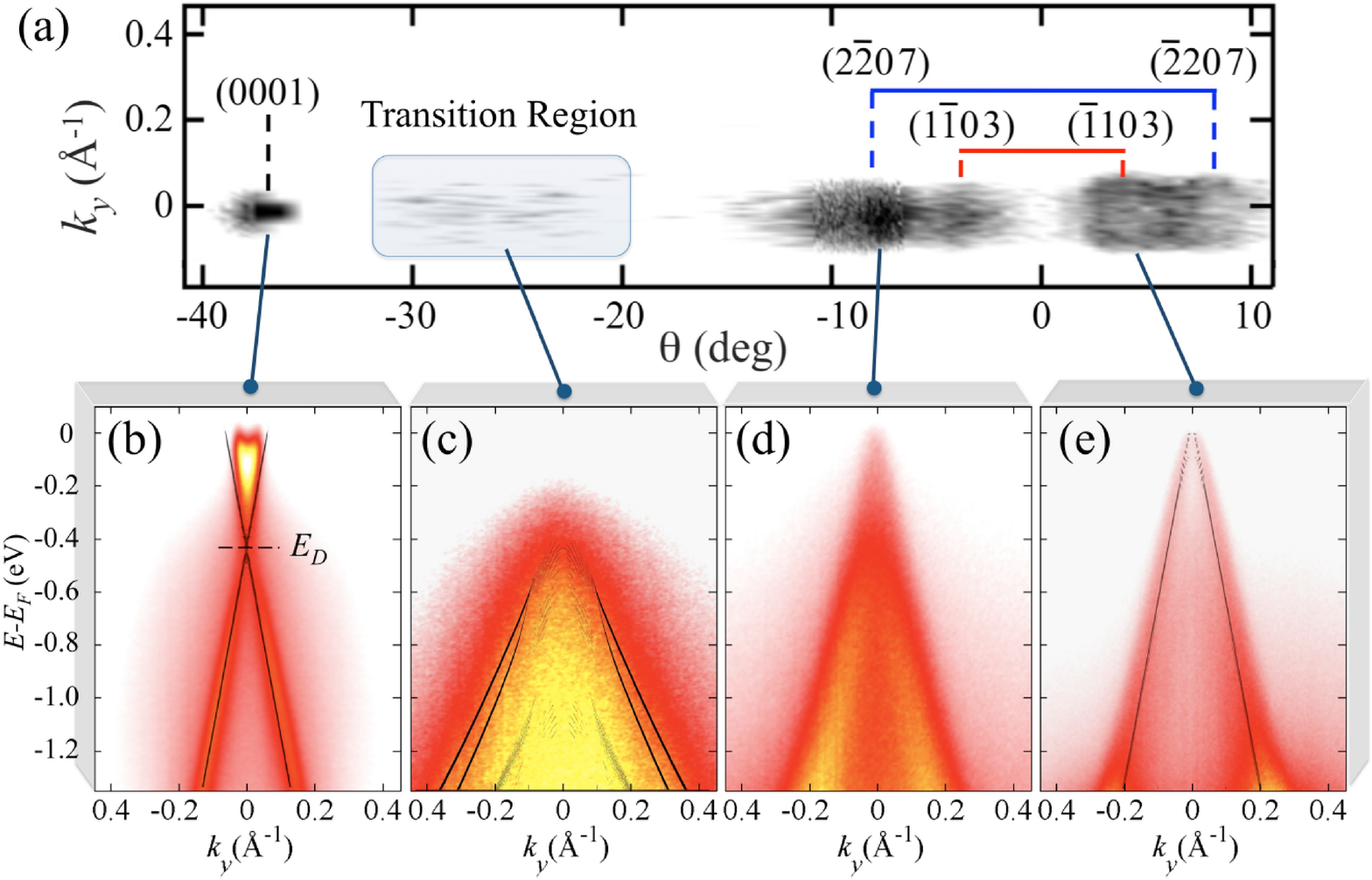}
\caption{(a) Quasi-Fermi surface from a 12 nm deep trench array after graphitization with the graphene ribbons' armchair edges parallel to the step edge.  The plot is a constant energy surface ($E\!-\!E_F\!=\!-0.43$ eV) in $k_y$ and tilt angle $\theta$ ($k_y$ is parallel to the graphene armchair edge [see Fig.~\ref{Ribbons}(d)]).  Dirac cones from each of the angular regions are shown in (b)-(e). (b) Dirac cone from the flat (0001) surface.  The spectrum is typical of a 1-layer graphene film above a graphene buffer layer.  The electron doping is 0.43 eV. (c) Typical distorted cone from the intermediate angular range shown in the grey box in (a).  (d)  Dirac cone from the $(2\bar{2}07)$ orientation. (e) Dirac cone from the $(1\bar{1}03)$ orientation. Lines in (b), (c), and (e) are the TB calculated bands for a 1.5 nm wide graphene strip between two doped graphene sheets [see supplemental material]. }\label{F:Cones}
\end{figure*}

\vspace*{2ex}
\noindent \textbf{Nano-patterened graphene ribbon arrays}

\noindent Epitaxial graphene on SiC is critical for these studies because it allows atomic control of where  graphene grows and what crystallographic orientation the graphene has relative to patterned features on the SiC.  On the SiC(0001) surface, graphene always grows rotated $30^\circ$ relative to the SiC $<\!10\bar{1}0\!>$ direction.\cite{Hass_JPhyCM_08} Using this directionality, SiC trenches where etched so that the growing graphene had its ``armchair'' edge parallel to the patterned steps. This allows the graphene Brillouin zone (BZ) orientation to be accurately specified with respect to the substrate so that ARPES can measure graphene's linear $\pi$-bands (Dirac cones) on both the (0001) surface and on facets formed from the sides of the pattern trenches.

During growth, the SiC trench sides facet into a stable new surface (see below). Graphene is known to flow from the (0001) surface, over the step, and onto these facets.\cite{Norimatsu_PhysicaE_10}  We have set the growth time and temperature to produce a single graphene layer above the ``buffer'' graphene-like sheet on the Si-face.\cite{Hicks_JPHYSD_12}  Both the buffer and monolayer graphene on the Si-face grow over the trench edge onto the sidewall facet and then bend back onto the (0001) face of the trench floor.\cite{Norimatsu_PhysicaE_10}  Figure \ref{Ribbons}(a) shows a schematic of the experimental ribbon geometry. The graphene that grows on the facets are referred to as sidewall graphene ribbons. An example of part of the post-graphitized arrays is shown in Figs.~\ref{Ribbons} (b) and (c).  The ribbons are extremely straight over many microns with a trench-height RMS roughness less than one 4H-SiC unit cell (1 nm). The estimated width of a sidewall ribbon is $W=d/\sin\Delta\theta_i$, where $d$ is the trench depth and $\Delta\theta_i$ is the angle between the surface normals of the sidewall facet and the (0001) face.  It is this long range order that allows detailed ARPES measurements

\vspace*{1ex}
\noindent \textbf{Topological changes in graphene's band structure}

 \noindent We have used ARPES to measure the topological dependence of graphene's electronic structure as it flows from the (0001) surface to the sidewall facets.  We do this by orienting the SiC steps edges with the $\theta$ rotation of the ARPES detector as shown in Fig.~\ref{Ribbons}(a).  In this geometry, with the graphene's armchair edge parallel to the SiC step edge, the detector measures $k_y$ (perpendicular to the $\overline{\Gamma K}$ direction) while $\theta$ scans $k_x(\theta,\hat{n}_i)$ along the $\overline{\Gamma K}$ direction, perpendicular to the step edges [see Fig.~\ref{Ribbons}(d)]. This experimental setup means that we measure graphene's $K$-point band structure for all orientations the graphene may take as it curves from the (0001) surface down over the trench walls.  If a Dirac cone appears in the detector at some $\theta_i$, we can identify its facet  normal $\hat{n}_i$ from the relative $\Delta\theta$ away from the (0001) $K$-point angle $\theta_o$.

Figure \ref{F:Cones}(a) shows a $\theta$ scan at constant energy from a 12 nm deep trench array. There is no intensity and therefore no cones between angles $-36^\circ$ to $-33^\circ$ and $-20^\circ$ to $-12^\circ$, indicating there are no stable facets in this range or that the facet size is extremely small. There are, however, four regions where cones appear.  Figures \ref{F:Cones}(b)-(e) show the ARPES measured cones from these regions.  The $K$-point Dirac cone at $\theta_o=-37^\circ$ is from the graphene on the (0001) flat surface [see Fig.~\ref{F:Cones}(b)].  This is the typical band structure of a single graphene layer above a buffer-graphene layer.\cite{Hass_JPhyCM_08} This graphene layer is n-doped by  $E_D-E_F=-0.43$ eV, which is the typical doping of a clean, ordered Si-face epitaxial graphene film.\cite{Riedl_JPhysD_10}

At small $\theta$ there are two sets of two Dirac cones with $\theta_i\!=\!\pm9^\circ \text{ and }\pm4.5^\circ$.   These cones correspond to graphene that has grown on $(2\bar{2}07)$, and $(1\bar{1}03)$ facets on one trench side and their complimentary $(\bar{2}207)$, and $(\bar{1}103)$ facets that form on the opposing trench side. These large area SiC facets developed and are stabilized during the graphitization process. The Dirac cones on these facets are slightly p-doped as discussed later. The sidewall graphene on these two facets are extremely well ordered. The $\Delta k_y$ width of these cones ranges from $0.04\text{\AA}^{-1}$ for 36 nm wide ribbons (the same width measured for an infinite graphene sheet on the (0001) surface) to $0.08\text{\AA}^{-1}$ for 15 nm wide ribbons.  Assuming that the $\Delta k$ width is entirely due to angular variations of the facet surface normal caused by step edge meandering, the RMS surface normal variation, over $1\mu$m in the $y$- and $x$- direction, is $\Delta\phi\!\sim\!2^\circ$ and $\Delta\theta\!\sim\!1.6^\circ$, respectively. Keeping in mind that these cones are a measured averaged over $>\!500$ parallel ribbons, the facet stability and long range order is exceptionally high. 
 
 \begin{figure}
\includegraphics[angle=0,width=8.0cm,clip]{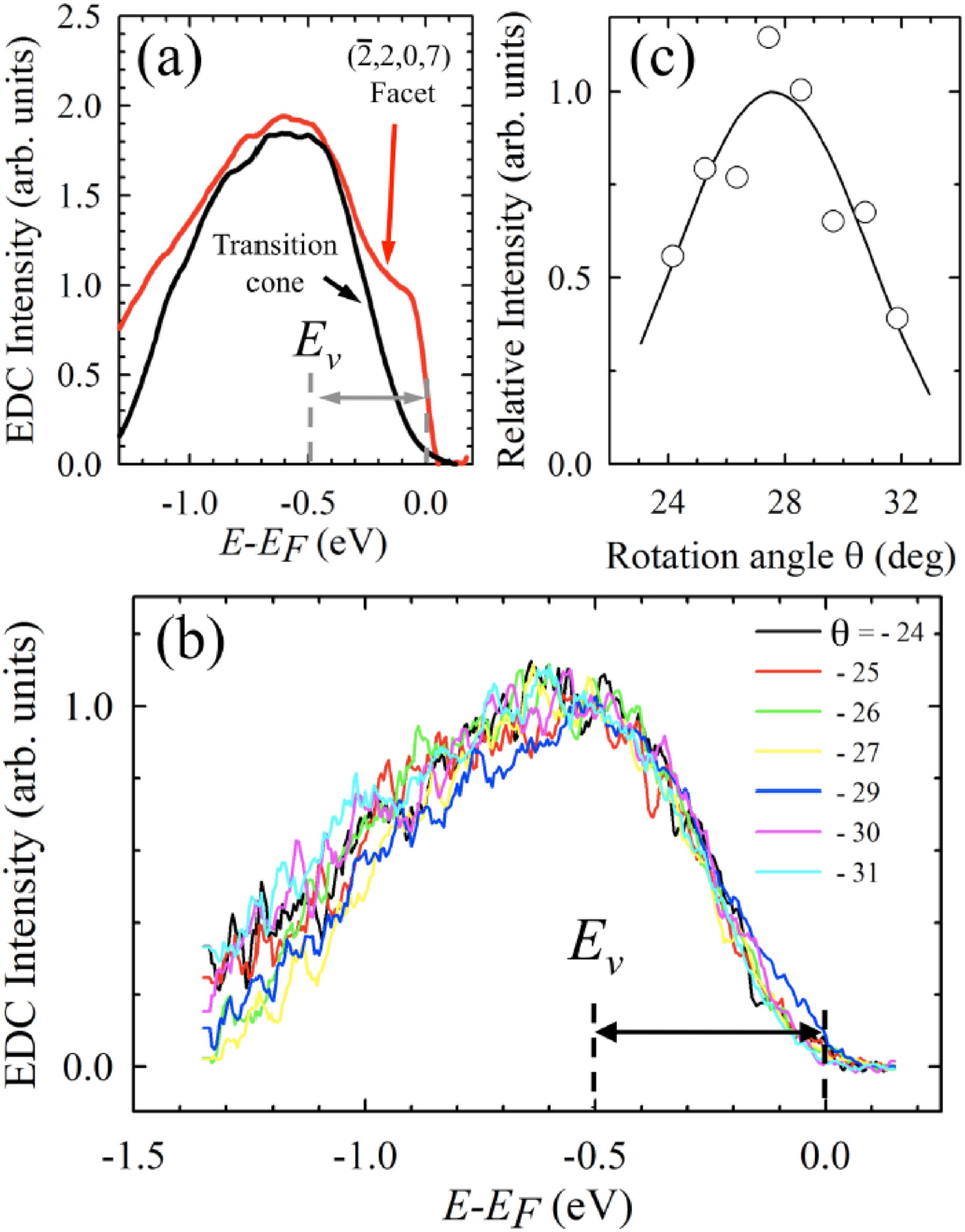}
\caption{(a) Comparison of the EDCs from the flat $(\bar{2}207)$ Dirac cone and a cone from the transition region (a linear background has been subtracted for clarity). EDCs are through $k_y\!=\!0$ for a 24 nm wide facet graphene ribbon.  The transition region cones shows that the graphene valance band is shifted 0.5 eV below $E_F$. 
(b)  EDCs of the cones in the transition region. Normalized scans are for different sample rotations, $\theta$. All cones in the transition region have the same profile, demonstrating the 1D nature of the region. (c)  Relative intensity of the transition cones as a function of $\theta$.  Line is a fit to the diffraction limit broadening of 30 eV photoelectron from a 1.4 nm wide source.}\label{F:Dope_Cones}
\end{figure}

The most important feature in Fig.~\ref{F:Cones}(a) is the transition between the flat 2-D graphene sheets on the (0001) and facet surfaces. This transition is evident in an angular range between $\theta\!=\!-33^\circ\!\le\!\theta\!\le-20^\circ$ where a dramatic change occurs in graphene's band structure, switching from discrete facet Dirac cones to a continuous distribution (in $\theta$) of broad diffuse cones. An example cone is shown in Fig.~\ref{F:Cones}(c). It is clear that this cones has a gap and a significantly lower slope (velocity $\propto\!\partial E/ {\partial k_y}$) than the Fermi velocity, $v_F$, measured for both the (0001) and facet Dirac cones.  Figure \ref{F:Dope_Cones}(a) compares an energy distribution curve (EDC) (intensity versus energy at a constant $k_y$) through the $K$-points for a transition region cone and the $(\bar{2}207)$ facet cone.  The spectra density for the transition cone decays rapidly below the valance band edge, $E_v$,  at $E\!-\!E_F\!\sim\!-0.5$ eV.  This is not a doping shift since there is no indication of $\pi$ bands above -0.5 eV.  Because the position of $E_F$ in the gap is not known, the lower limit on the gap size is 0.5 eV.  As we'll now show, this gapped region is a manifestation of a nearly one-dimensional semiconductor.
 
\vspace*{1ex}
\noindent \textbf{Junction Potential Boundary }
  \noindent The gapped graphene region between the (0001) surface and the sidewall facet is unique because it represents a nearly 1-D graphene semiconductor seamlessly connected to two 2-D graphene sheets on its sides. The nearly 1-D electronic nature of the bent graphene can best be demonstrated by considering what a hypothetical ARPES experiment would measure in a better known 1-D system, a carbon nanotube (CNT). If we had enough photon flux to make an ARPES measurement on a single CNT with its axis aligned in the $y$-direction, we would only see dispersion along $k_y$.  All other $k$'s are not good quantum numbers because of the 1-D nature of the CNT.  In other words, the APRES spectrum would be the same no matter what $\theta$ we rotated the CNT about the $y$-axis, i.e the $k_y$ dispersion will be nearly independent of $\theta$. 

If we think of the curved graphene region as a partial arc of a CNT, then the thought experiment is in fact identical to the experiment in this work.  The only difference is that we use 300-500 self-assembled parallel coherent ribbons to give sufficient intensity to make the measurement possible. Because the step edges are parallel to the $\theta$ rotation axis, we rotate around the axis of the bent graphene by $\theta$.  The results for the ``edge-CNT'' are shown in Fig.~\ref{F:Dope_Cones}(b).  Over a $12^\circ$ rotation the dispersion $E(k_y)$ is independent of $\theta$, i.e, $E$ is independent of $k_x(\theta)$.  This simple experiment demonstrates that the gapped graphene is confined to a very narrow region of the film. 

Unlike the CNT the curved edge is a finite strip.  This means that the angular distribution of the ARPES is not completely independent of $\theta$. The measured intensity is modulated by a $\theta$-dependent envelope function of a finite size object.   We can use this envelope function to estimate the coherent dimension of the semiconducting bend.  The  $\Delta\theta\sim\!12^\circ$ angular range of the gapped transition spectra can be viewed as the diffraction limit broadening of a 30 eV kinetic energy photoelectrons electron ($\lambda\!=\!0.22$ nm) from a source with dimension $d$. Figure \ref{F:Dope_Cones}(c) shows the relative intensity of a cone as a function of $\theta$ in the transition region. A fit of the measured $I(\theta)$ to the broadening function $\text{sinc}^2(\pi d\text{sin}\theta/\lambda)$ [see Fig.~\ref{F:Dope_Cones}(c)] gives $d\!=\!1.4$ nm. This width is consistent with transmission electron microscopy, TEM, measurements that show graphene bending over step edges in 5-10 graphene lattice constants.\cite{Norimatsu_PhysicaE_10} While local probes like STM will be able to measure this width more precisely, it is clear that the semiconducting graphene strip is very narrow.

Of course, the bent graphene band structure cannot be understood in the context of a CNT since the wave function boundary conditions are different in the two system. Instead of cyclical boundary conditions in a complete CNT, the bent graphene can be considered as a finite arc that is continuously bounded by two flat graphene sheets on both sides with different doping.  While a self consistent theory will need to developed, there are several important effects that will influence the band structure of this system; strain, finite size effects, local coulomb potential, and substrate bonding.  

The strain and electronic energies are comparable in this system. From Norimatsu and Kusunoki's work, the known radius of curvature of a graphene layer over a step is $\sim\!1\text{nm}$.\cite{Norimatsu_PhysicaE_10}  The strain associated with this amount of bending can be very large.  Calculated band-gaps for similarly strained graphene can range from 0.2-0.5 eV depending on the model.\cite{Ni_ACSN_08,Pereira_PRB_09,Wong_JPC_12}  Comparing this strain energy with a finite size band-gap expected for a 1.4 nm graphene ribbon, $E_g\sim\!1\text{eV-nm}/1.4\text{nm}\!=\!0.8$ eV,\cite{Wakabayashi_STAM_10,Son_PRL_06}  shows that both effects lead to band gaps similar to the experimental value.  Because the energy scales are similar, Peierl's like distortion are potentially important.  This is an important point because, unlike a CNT or a finite width ribbon, the 1-D strip's boundary conditions are not discrete.  This allows the graphene at the bend to flex in order to lower its total energy.
 
Besides the effect of strain and boundary conditions, there is an onsite potential at the bend due to the different doping on either side of the ribbon.  The charge distribution and the Schottky barriers that can form at the metal-semiconducting will require a self consistent solution to the Poisson equation.  This charge distribution affects the electron boundary condition and will be important in determining resonant states that influence transport through the metal-semiconductor-metal region.

 Because the Schottky barrier is so narrow, tunneling through the junction should be an important part of graphene transport over steps. Evidence for this, in the form of a decrease in graphene's conductivity over randomly oriented 1 nm SiC steps, has been reported.\cite{Ji_NM_12}  Although graphene's strain going over a 1 nm step should be significantly smaller than the taller steps studied in this work, the transport findings in Ref.~[\onlinecite {Ji_NM_12}] are consistent with a semiconducting strip at the step.  
 
  While it is obvious that a good theoretical understanding of the electronic structure remain to be formulated, we have used a simple tight binding calculation that reproduces many of the salient features of the ARPES data [see supplement].  The model consists of a 2000 atom wide flat graphene sheet. A 1.5 nm strip in the middle of the 2D film has a different hopping integral than the rest of the sheet to mimic the effect of strain at the bending.\cite{Chamon_PRB_00,Pereira_PRB_09,Ribeiro_NJP_09,NetoRMP81-109-2009}  In addition, a position dependent on-site potential, $V(x)$, is used.  This is to account for the transition from the n-doped (0001) graphene and the p-doped sidewall graphene [see below and in the supplemental material]. The band structure calculated using this model is shown in Figs.~\ref{F:Cones}(b), (c), and (e).  Note that the model produces a significant number of state within the cones from the narrow semiconducting region.  This correlates well with the experimental diffuse intensity observed within the cone. 
  
   \begin{figure}
\includegraphics[angle=0,width=8.0cm,clip]{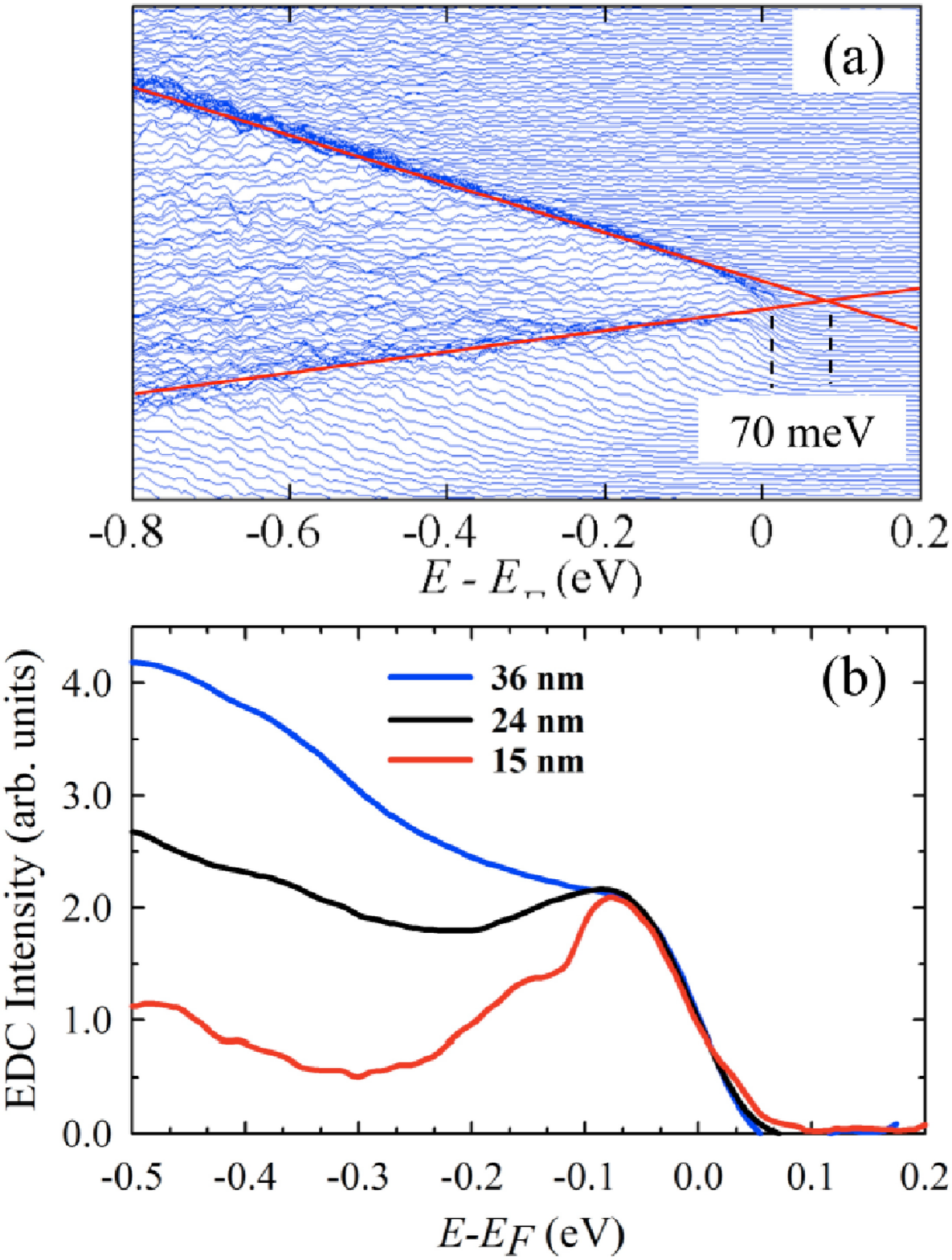}
\caption{(a) A magnified view of the graphene Dirac cone from the stable $(1\bar{1}03)$ facet Fig.~\ref{F:Cones}(e) near $E_F$.  The figure shows EDC offset in intensity to highlight the region of the cone near $E_F$. The facet is p-doped by 70 meV$\pm20$ meV. (b) EDCs through $k_y=0$ of the $(1\bar{1}03)$ facet cones.  Data is for three different sidewall ribbon widths; 15, 24, and 36 nm.}\label{F:1D_Region}
\end{figure}

\vspace*{2ex}
\noindent \textbf{Facet band structure}

 \noindent  The morphology shown in Fig.~\ref{Ribbons}(a) consists of a wide band gap, nearly 1-D, semiconducting wire bounded by two flat graphene sheets. One sheet on the (0001) surface and the other on the sidewall facet. Because boundary conditions will be important in understanding the details of the ribbon band structur, we must understand the differences between the graphene grown on the (0001) surface and graphene grown on the sidewall facets.  The two sheets each are in different charge states.  Graphene on the (0001) surface is metallic and n-doped. The sidewall facet graphene has a small p-doping. Figure \ref{F:1D_Region}(a) shows that the $(1\bar{1}03)$ facet has a $\sim\!70$ meV p-doping.  It is not clear from these studies if the narrow sidewall ribbons, have a gap, particularly the 15 nm wide ribbons.  Figure \ref{F:1D_Region}(b) shows EDCs through the graphene Dirac cone on the $(1\bar{1}03)$ facet for different ribbon widths.  All ribbons show the same shape near $E_F$.  The expected gap for a $w=15$ nm ribbon is $E_g\sim 1\text{eV-nm}/15 \text{nm}=66$ meV.\cite{Wakabayashi_STAM_10}  Since the doping is comparable to the expected gap, it is difficult to say with certainty whether or not the facet graphene is metallic or semiconducting.  We can say that if there is a gap, it is less than 70 meV.

Why there is a doping difference between graphene on the (0001) and graphene on either the $(2\bar{2}07)$ or  $(1\bar{1}0 3)$ SiC surfaces is worth attention because it suggests other ways to spatially control graphene's properties. The surface reconstructions are different on the two facets. LEED shows that while the graphene diffraction spots are visible on the stable facet walls, the \sixrt reconstruction spots of the (0001) face disappear [see supplemental material].  The surface reconstruction must be different on the two facets because the bulk terminated sidewall facets have a rectangular symmetry with two dangling bonds per surface atom while the (0001) surface has hexagonal symmetry with one dangling bond per surface atom.  The different surface symmetries implies that polarizability, bond re-hybridization, chemical reactivity, etc., can all be different and influence the charge transfer.  While thorough studies of these facets remain to be done, we can show that the work function for the sidewall facet is very different from that on the (0001) surface. 

\begin{figure}
\includegraphics[angle=0,width=7.0cm,clip]{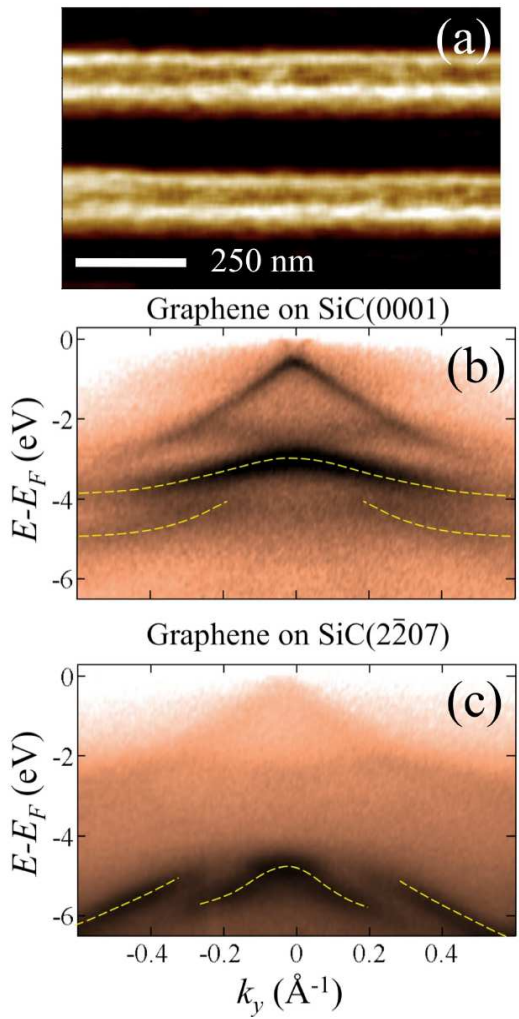}
\caption{ (a) An EFM image of the side wall graphene with 36 nm wide facets. Light lines are areas were the work function changes when graphene has grown on the sidewalls. (b) and (c) are valence band spectrum near the $K$-point for the flat (0001) and $(2\bar{2}07)$ facet surfaces, respectively.  The dashed line shows the surface derived band change.} \label{TiltBZ}
\end{figure}

A closeup electric force microscope (EFM) image of the ribbons in Fig.~\ref{TiltBZ}(a) shows the work function contrast change on the sidewalls that occurs after graphene has grown on the sidewalls.  From the photoelectron secondary electron cut off, we have measured a $\sim\!1.5$ eV decrease in the work function of the facet relative to the polar (0001) face. This implies that the buffer graphene layer on the (0001) surface reacts very differently than the graphene layer above the SiC sidewall facet. This is confirmed in Figs.~\ref{TiltBZ}(b) and (c) that show a surface-derived\cite{Emtsev_PRB_08} set of bands near the $K$-point that both shift to higher binding energy and change shape on the facet.  We note that the shift down is nearly equal to the work function shift even though the Dirac cones remain at the same energy.  In some sense the reconstruction of the facet, with a presumed interaction change with the graphene buffer layer, behaves similar to the hydrogen intercalated SiC(0001)-graphene system,\cite{Riedl_PRL_09} where the hydrogen affects the buffer-SiC interface in a way that effectively neutralizes the charge transfer to the graphene film. 

One last point must be mentioned. The bilayer graphene pair on the large facets should have an unusual stacking.  Assuming that the two sheets bend over $29^\circ$  (the angular distance to the first flat facet), the top graphene layer must slip relative to the second layer by $\sim30\%$ of the graphene unit cell.  The strain energy to keep the two layers in an \textit{AB} commensurate stacking with this amount of slip is too large and cross-sectional TEM shows no evidence for buckling or any other strain relieving mechanism that would be required to bring the two sheets into commensuration as they go over the step edge. It has recently been shown that sliding graphene layers can significantly distort the $K$-point band structure.\cite{sliding_graphene}  This may explain while the experimental two layer Dirac cones are similar to those from a single graphene sheet and do not show the \textit{AB} split bands associated with Bernal stacking.  A closer inspection of the cones from the large stable facets does shows a large spectral intensity $\sim\!0.5$ eV below $E_F$ [see Fig.~\ref{F:Dope_Cones}(a)] that would be characteristic of a disorder induced energy-broadening of the split band from an \textit{AB} stacked pair.  However, no significant band broadening in $k_y$, which must be associated with the \textit{AB} band, is observed.
      
\vspace*{2ex}
\noindent \textbf{Conclusion}

\noindent  We have shown that a nearly 1-D, large band-gap, semiconducting graphene ribbon, seamlessly connected to metallic graphene sheets, can be produced from a scalable bottom-up approach by forcing topological changes in the SiC that in return affect graphene's electronic properties.  The ribbons produced this way have a band-gap energy $>\!0.5$ eV and are both narrow ($\sim\!1.4$ nm) and continuous over macroscopic lengths.  In a sense, the macroscopic arrays of graphene metallic-semiconducting junctions presented in this work are the ideal architecture envisioned for carbon nanotube electronics;  perfectly aligned ribbons with identical chirality patterned over macroscopic distances with atomic precision.  The rough edge problem associated with lithographically patterning graphene ribbons is solved using a the bottom-up assembly as the graphene grows.\cite{Sprinkle_NN_10}  The ``edges" in the structures presented here are atomically perfect because they are simply a curved distortion in a continuous graphene sheet that is defined by potential gradients (elastic or electronic) rather than ill defined edge terminations.

Graphene, and therefore graphene electronics, is too often viewed as an isolated entity. In this work we do not merely show the importance of the graphene-substrate interaction,  we demonstrate how the support substrate can be used to purposely alter graphene's electronic properties.  We give an important example of how a \emph{well-ordered}, \emph{commensurate} substrate interaction can be used to force topological changes in the graphene that lead to metal-semiconducting transitions and alter its doping. The large number of ways the substrate can be altered means that a new field of scalable geometry dependent device research is just beginning.  For instance, we have only studied armchair edge 1-D wires.  Like the chirality in CNTs the device properties could be very different for zigzag edge bends.  Another example is doping graphene.  In this work the facet doping is controlled by intrinsic SiC surface effects.  Because the facet walls have a different chemical reactivity, adsorbates can be designed with self-directed bonding to sites only present on the facet.  This means that graphene dopants can be patterned with atomic control simply by cutting trenches in different crystallographic edges. Exploring the transport properties of the junctions described in this work is just beginning.  New architectures based on the principles outlined here are sure to significantly alter graphene electronics research.

\begin{acknowledgments}
We wish to thank A. Zangwill for many useful discussions. This research was supported by the W.M. Keck Foundation, the Partner University Fund from the Embassy of France and the NSF under Grant No. DMR-0820382 and DMR-1005880.  We also wish to acknowledge the SOLEIL synchrotron radiation facilities and the Cassiop\'{e}e beamline. J. Hicks also wishes to acknowledge support from the NSF-GRFP under Grant No. DGE-0644493.

\end{acknowledgments}

\vspace*{2ex}
\noindent \textbf{Methods}

 \noindent The substrates used in these studies were n-doped $n\!=\!2\times\!10^{18}\text{cm}^{-2}$ 4H-SiC. The graphene ribbons arrays were made by a selective growth method.\cite{Sprinkle_NN_10}  Vertical trenches are first produced in the SiC(0001) (Si-face) surface and then graphitized in a controlled silicon sublimation furnace.\cite{WaltPNAS} The arrays were prepared by first producing a negative mask of hydrogen silsesquioxane (HSQ) by e-beam lithography.  The SiC(0001) substrate was exposed to reactive ion etching with a $\text{SF}_6\!-\!\text{O}_2\!-\!\text{Ar}$ plasma. This produced 7.5 nm to 18 nm deep trenches depending on etching time.  The samples were then graphitized by heating to 1560\degC in a carbon RF furnace. The high density parallel trench arrays were made with a pitch between 100-400 nm over a $1\!\times\!3\text{mm}^2$ area (2,500-10,000 1mm long trenches).   This allows 300 to 500 ribbons to be easily localized in the $\sim\!40-50\mu$m ARPES beam.

  The samples were transported in air before introduction into the UHV analysis chamber.  Prior to ARPES measurements the graphene films were thermally annealed at 800\!\degC in UHV.   ARPES measurements were made at the Cassiop\'{e}e beamline at the SOLEIL synchrotron in Gifs/Yvette. The high resolution Cassiop\'{e}e beamline is equipped with a modified Petersen PG-monochromator with a resolution $E/\Delta E \simeq 70000$ at 100 eV and 25000 for lower energies. The detector is a $\pm 15^{\circ}$ acceptance Scienta R4000 detector with resolution $\Delta E\!<\!1$ meV and $\Delta k\!\sim\!0.01\text{\AA}^{-1}$ at $\hbar \omega\!=\!36$ eV. All measurements were carried out at 100K. The total measured instrument resolution is ($\Delta E\!<\!12$ meV).









\end{document}


\title{SUPPLEMENT\\
A wide band gap metal-semiconductor-metal nanostructure made entirely from graphene}

\author{J. Hicks}
\affiliation{The Georgia Institute of Technology, Atlanta, Georgia 30332-0430, USA}
\author{A. Tejeda}
\affiliation{Institut Jean Lamour, CNRS - Univ. de Nancy - UPV-Metz, 54506 Vandoeuvre les Nancy, France}
\affiliation{Synchrotron SOLEIL, L'Orme des Merisiers, Saint-Aubin, 91192 Gif sur Yvette, France}
\author{A. Taleb-Ibrahimi}
\affiliation{UR1 CNRS/Synchrotron SOLEIL, Saint-Aubin, 91192 Gif sur Yvette, France}
\author{M.S. Nevius}
\author{F. Wang}
\author{K. Shepperd}
\author{J. Palmer}
\affiliation{The Georgia Institute of Technology, Atlanta, Georgia 30332-0430, USA}
\author{F. Bertran}
\author{P. Le F\`{e}vre}
\affiliation{Synchrotron SOLEIL, L'Orme des Merisiers, Saint-Aubin, 91192 Gif sur Yvette, France}
\author{J. Kunc}
\author{W.A. de Heer}
\affiliation{The Georgia Institute of Technology, Atlanta, Georgia 30332-0430, USA}
\author{C. Berger}
\affiliation{The Georgia Institute of Technology, Atlanta, Georgia 30332-0430, USA}
\affiliation{CNRS/Institut N\'{e}el, BP166, 38042 Grenoble, France}
\author{E.H. Conrad}\email[email: ]{edward.conrad@physics.gatech.edu}
\affiliation{The Georgia Institute of Technology, Atlanta, Georgia 30332-0430, USA}


\maketitle


\vspace*{2ex}
\noindent \textbf{LEED from patterned arrays}

\begin{figure}[htb]
\includegraphics[angle=0,width=8.0cm,clip]{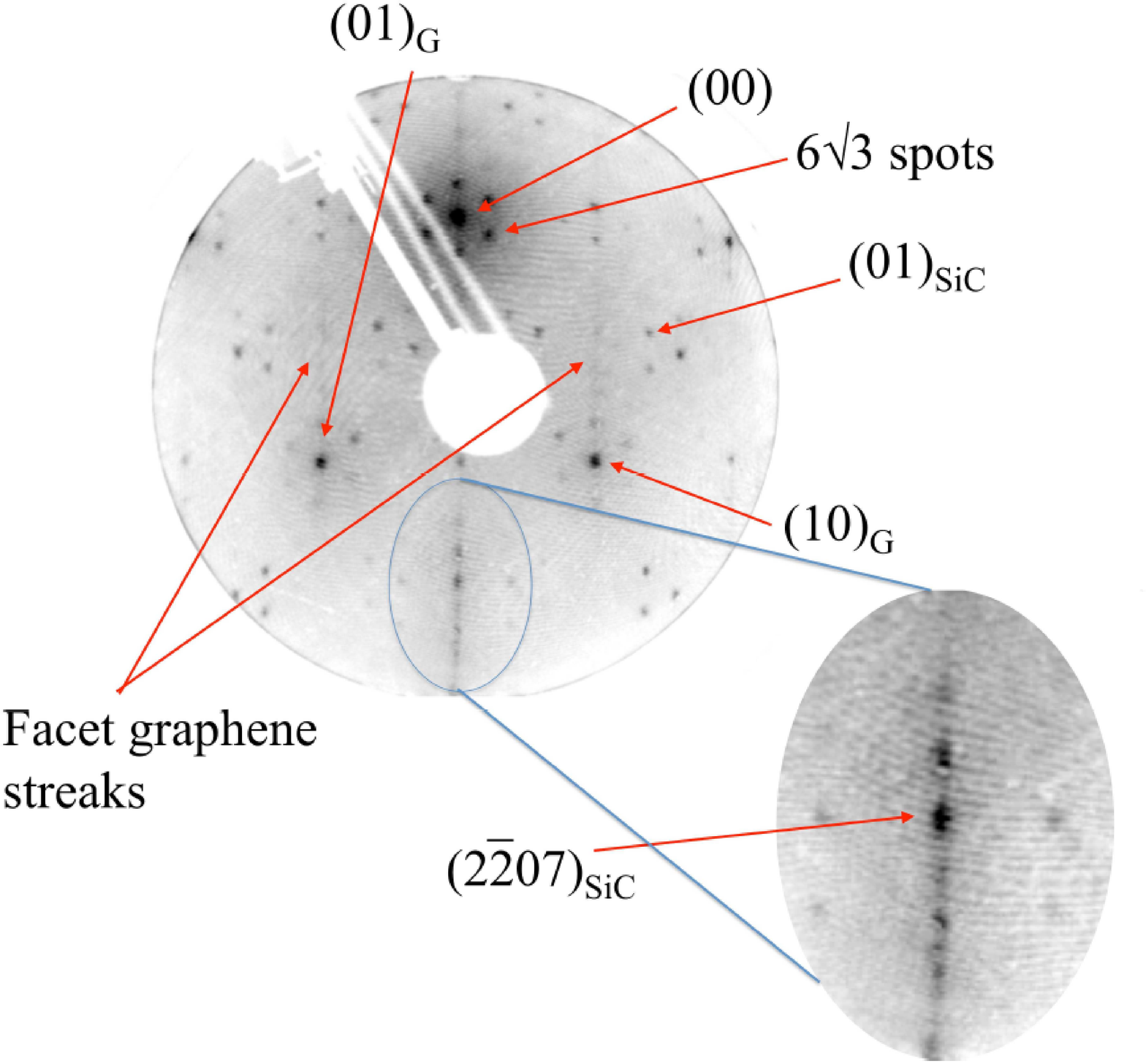}
\caption{(a) LEED pattern from a graphitized trench array with side wall graphene.  The trench depth is 20 nm and the trench pitch is 100 nm. The incident energy is 65 eV.  The the spots index with subscript "G" are graphene spots oriented $30^\circ$ relative to the SiC(0001) surface.  The $6\sqrt{3}$ and the $(01)_\text{SiC}$ spots from the (0001) surface are also marked.  Intensity streaks run through the graphene and (00) rods.  Inset shows the specular rod from the $(2\bar{2}07)$ facet.}\label{F:LEED}
\end{figure}

\noindent Because the ribbon arrays are highly ordered, it is possible to get meaningful information about the facet structure from LEED.  Figure \ref{F:LEED} shows a LEED pattern from parallel ribbon arrays with 20 nm trench depth.  The main features of the pattern are the graphene rods from the (0001) surface with the typical $6\sqrt{3}$ pattern.\cite{Hass_JPhyCM_08} The graphene is rotated $30^\circ$ with respect to the SiC  $<\!01\bar{1}0\!>$ direction. The trench edges are oriented perpendicular to the SiC $<\!01\bar{1}0\!>$ direction so the graphene grows over the trench edges with the graphene $<\!110\!>_\text{G}$ direction perpendicular to the steps.  At the electron energy used in the figure, the SiC spots are reduced in intensity both from a structure factor effect and the electron attenuation through the graphene.

Note the streaking of the $(00)$, $(01)_\text{G}$, and $(10)_\text{G}$ spots.  The streaking is typical of a faceted surface.\cite{Conrad_LEED} It is important to realize that only the graphene spots are streaked.  This conclusively shows that the graphene is on the sidewalls.  The $6\sqrt{3}$ spots show no streaking as expected because this pattern is only associated with the (0001) surface graphene.

Because the position of the specular spot (momentum parallel to the surface is zero) is independent of energy, the specular spot from the facet surface can be identified by changing the electron energy and finding the spot whose position is independent of energy. From such a series of LEED images, we can identify the specular spot from graphene on a SiC facet at an angle of $\sim\!29^\circ$ relative to the (00) rod, i.e, the $(2\bar{2}07)$ facet.  This marked in the insert in Fig.~\ref{F:LEED}.

\vspace*{2ex}
\noindent \textbf{Tight binding calculations}

\begin{figure}[h!]
	\centering
		\includegraphics[width=14cm]{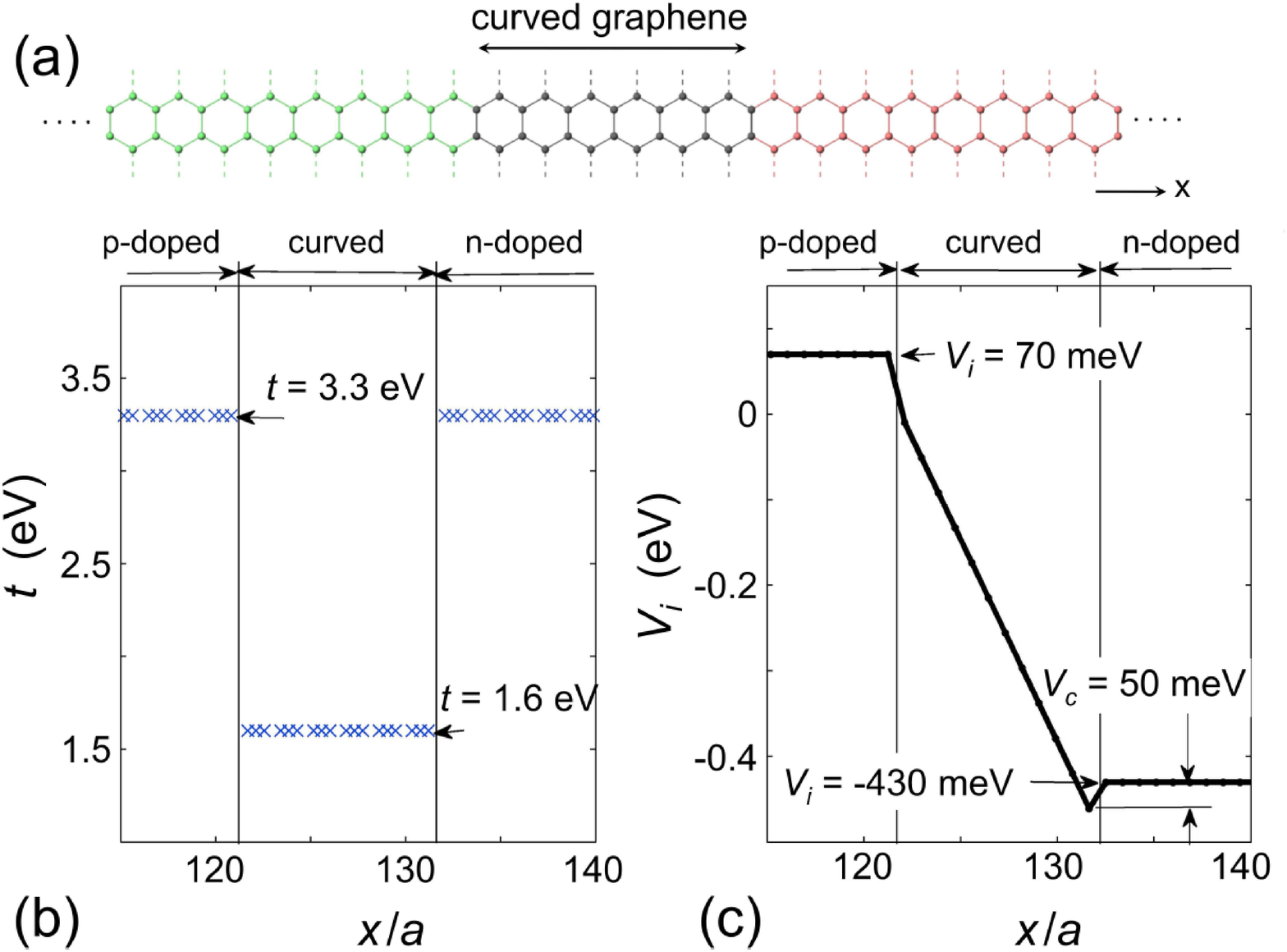}
		\caption{(a) A model of the metallic-semiconducting-metallic graphene ribbon.  The p-doped part spans $x=0-17.25$~nm, curved part $x=17.25-18.75$~nm and n-doped part $x=18.75-122.9$~nm. The variation of (b) the hopping parameter $t(x)$ and (c) the on-site potential $V_i(x)$ across the GNR.  We use 1000-GNR ($N=1000$). The length is in the units of graphene C-C bond $a=1.42$~$\mathrm{\AA}$. The hopping parameters in different parts of the ribbon were determined from the measured slope of $E(k)$ dispersion. The on-site potentials describe the n- ($V_{i,n}=-0.43$~eV) and p- ($V_{i,p}=+0.07$~eV) doping, smooth potential variation in the curved part of the ribbon and an additional on-site potential ($V_c=50$~meV) induced by the curvature $R=3$~nm~\cite{NetoRMP81-109-2009}.}
		\label{fig300AGNR}
\end{figure}

\noindent An estimate of the band structure of the 1-D bent graphene, continuously connected between two metallic graphene sheets with different doping, was calculated using a tight binding method in the nearest-neighbor approximation.  We model the system using a 1.5 nm flat graphene nano-ribbon, GNR, with armchair edges to represent the bent graphene.  This strip is connected to a 17.25 nm wide p-doped strip representing the side wall graphene and an n-doped 104.15 nm wide strip representing graphene on the (0001) surface [See Fig.~\ref{fig300AGNR}(a)]. The total number of carbon atoms across the three regions is 2000. The model is periodic in the $y$-direction, perpendicular to the armchair edge.

For the purposes of demonstrating the confining effect of strain in the bend, we assuming a position dependent hoping parameter, $t(x)$.\cite{Chamon_PRB_00,Pereira_PRB_09,Ribeiro_NJP_09} We allow $t(x)$ to be different in the narrow ribbon compared to the large graphene sheets on either side.  To include the different doping of the three graphene sheets, we allow a spatially varying on-site potential $V_i(x)$. The hamiltonian describing the system is then;

\begin{equation}
	\hat{H}=-t(x)\sum_{\langle i,j\rangle}\hat{a}_i^+\hat{b}_j+\sum_{i}V_i(x)\hat{n}_i+h.c.,
\end{equation}

where $\hat{a}_i$ ($\hat{a}_i^+$) and $\hat{b}_j$ ($\hat{b}_j^+$) are the annihilation (creation) operators for the sublattice $A$ and $B$, respectively. The operator $\hat{n}_i$ is the local charge density operator~\cite{NetoRMP81-109-2009} and is given by $\hat{n}_i=\hat{a}_i^+\hat{a}_i$ ($\hat{n}_i=\hat{b}_i^+\hat{b}_i$) for the $A$ and $B$ sublattice. Figures \ref{fig300AGNR}(b) and (c) show the spatial variation of $t(x)$ and $V_i(x)$, respectively.  Because calculating $V_i(x)$ in the ribbon is beyond the scope of this calculation, we describe the potential in the ribbon with a 50 meV deep well, smoothly connected to the two differently doped border graphene sheets as shown in Fig.~\ref{fig300AGNR}.  While the shape of this well influences the resonant states in the system, it does not change the important differences of the band structure in the three regions.

We solve the eigenvalue problem $\hat{H}\psi_{\nu,k_x}=E_{\nu,k_x}(k_y)\psi_{\nu,k_x}$ in the armchair nano-ribbon geometry using $2N=2000$ carbon atoms. This leads to 2000 bands, $E(k_y)$, that are indexed by the quantum numbers $\nu$ and $k_x$. The quantum number $\nu$ indexes the sub-bands in the conduction and valence bands. $k_x$ is the wave number of the standing waves in the narrow ribbons (1D limit) and becomes the Bloch wave number of the propagating states in the limit of infinitely wide ribbons (2D limit where GNR becomes a graphene sheet). 

In order to compare the results of this model to the ARPES data, we have to take into account the selectivity of the ARPES with respect to $k_x$ [see Fig.~1 in the main text]. To do this, we must project the calculated states $\psi_{\nu,k_x}$ onto the states $|k_{x,exp}>$, where $k_{x,exp}$ is the the component of $k_x$ parallel to the local surface normal.  $k_{x,exp}$ is selected by the angle $\theta$ of the detector with respect to the (0001) facet. This is important for properly representing the band structure at the $K$-point of the two flat sheets.  This projection is not necessary on the curved part of the ribbon because the 1-D character of the ribbon. In that case, all the states $\psi_{\nu,k_x}$ for the narrow ribbon are mapped onto the 1D Brillouin zone given by $k_y=\left\langle -\frac{\pi}{3a},+\frac{\pi}{3a}\right\rangle$ so that the measured band structure is independent of $\theta$.